\documentclass[11pt]{article}
\usepackage{latexsym,graphicx,amssymb,color,amsmath}
\usepackage[width=450pt,height=675pt]{geometry}
\usepackage[bottom]{footmisc}
\usepackage[numbers,sort&compress]{natbib}
\usepackage{hyperref}
\usepackage{hypernat}
\usepackage{relsize}

\newcommand{\be}{\begin{equation}}
\newcommand{\ee}{\end{equation}}
\newcommand{\ba}{\begin{eqnarray}}
\newcommand{\ea}{\end{eqnarray}}

\begin{document}
\vspace*{-2cm}
\begin{flushright}
SPIN-08/09, ITP-UU-08/09, DCPT/08/09
\end{flushright}
\vspace{0ex}

\begin{center}
{\larger\larger\larger {\bf Dynamics of icosahedral viruses: what does\\[1ex] Viral
    Tiling Theory teach us?}}\\[2ex]
{\smaller\smaller (Contribution to the proceedings of the `Second Mathematical Virology Workshop', Edinburgh
      (6-10 August 2007)}\\[-1ex]
\vspace{1cm} {\large  \bf Kasper 
Peeters\,\footnote{\noindent E-mail: {\tt kasper.peeters@aei.mpg.de}} and Anne
Taormina\,\footnote{\noindent E-mail: {\tt anne.taormina@durham.ac.uk}}}\\

\vspace{0.3cm} {${}^{1}$}\em Institute for Theoretical Physics\\ Utrecht University\\
P.O. Box 80195, 3508 TD Utrecht, The Netherlands.\\
 \vspace{0.3cm} {${}^{2}$\em \it Department of Mathematical
Sciences\\ University of Durham\\ Durham DH1 3LE, U.K.}\\ 
\end{center}
\medskip

\begin{abstract}
We present a top-down approach to the study of the dynamics of
icosahedral virus capsids, in which each protein is approximated by
a point mass. Although this represents a rather crude coarse-graining,
we argue that it highlights several generic features of vibrational
spectra which have been overlooked so far.  We furthermore discuss the
consequences of approximate inversion symmetry as well as the role
played by Viral Tiling Theory in the study of virus capsid vibrations.
\noindent \end{abstract}
\vskip 1cm
\section{Introduction}

It has been experimentally observed that viruses can alter their shape
to fulfill specific functions. In particular, they may swell during
maturation~\cite{Canady:2000a,Conway:2001a,Wikoff:2000a,Heymann2003a,Kuhn:2002a,Jiang:2003a},
twist to release their genetic material during infection, or morph
during assembly. Such large scale conformational changes are
consistent with the widespread hypothesis that viruses do vibrate, and
it is therefore of interest to study their dynamics with the help of
mathematical and computational techniques which have been tried and tested in the context of biomacromolecule vibrations (see~\cite{Chou:1988} for a review).

Normal mode analysis is one such
method~\cite{Noguti:1982a,Noguti:1983a,Brooks:1983a,Levy:1984a}, which
has been successfully applied to the study of proteins and a variety
of viruses to
date~\cite{Jernigan:2003,Tama:2002a,Tama:2005a,Vlijmen:2005a}. A major
challenge is the huge number of degrees of freedom involved in such
systems. Several degrees of coarse-graining, as well as group
theoretical methods (inspired by their successful application to small
molecules and fullerenes~\cite{Wilson:1934, Weeks:1989a,Wu:1987a}),
have been implemented in computer simulations in order to extract
information on the low-frequency modes of vibration which are thought
to be relevant for protein and virus
function~\cite{Simonson:1992,Vlijmen:2001}.  Although such theoretical
data become increasingly available for icosahedral
systems~\cite{Vlijmen:2005a,Tama:2005a, Freddolino:2006a} thanks to
advances in computer power, a clear and insightful vibrational pattern
across icosahedral viruses has not emerged yet. The art of
coarse-graining is a delicate one, as it is often argued that
excessive coarse-graining produces a dynamical picture that has little
to do with reality. We actually need a hierarchy of coarse-grained
calculations, which hopefully reveal complementary aspects of the
dynamical jigsaw.
\medskip

We argue here that even the crudest approximation, where each capsid
protein is treated as a point mass located at its centre of mass, is
helpful in highlighting dynamical features that are present in more
sophisticated normal mode analyses, but have been overlooked so
far. Our initial mathematical motivation was to assess to which extent
Viral Tiling Theory, a recently proposed model for icosahedral viral
capsids which solves a classification puzzle in the Caspar-Klug
nomenclature \cite{Twarock:2004a,Twarock:2005b}, provides a new insight in the dynamics of
viruses. In particular, we ask whether there is a correlation between
the vibrational patterns of viruses with a given number of coat
proteins and their viral tiling.

The paper is organised as follows. In Section 2, we briefly describe
Viral Tiling Theory in the context of the viral capsids RYMV ($T=3$),
HK97 ($T=7\ell$) and SV40 (pseudo $T=7d$), with emphasis on how the
underlying icosahedral symmetry manifests itself in different subtle
ways for these three cases. In particular, it has implications for the
group theoretical analysis of normal modes of vibrations. An expanded
version of these remarks, applicable to viruses and phages of all $T$
numbers, is available in \cite{ElSawy:2007a}. Section 3 provides a simple
normal mode analysis for the three capsids above, where group
theoretical techniques reminiscent of those used in calculations of
vibrational modes of small molecules are implemented. This paves the
way for the more extensive study performed in \cite{kas_virdyn}, which
reveals an intriguing universal pattern of low frequency normal
modes. We conclude with some open questions prompted by our
investigations.

\section{Tilings of Rice Yellow Mottle, Hong-Kong 97 and Simian Virus 40}

Viral tiling theory provides an elegant way of encoding the
icosahedral symmetry of viral capsids by keeping track of the location
of coat proteins and the orientation of capsomers on the viral shell,
while also keying in the dominant\footnote{In some cases, there exist
  stronger bonds between proteins pertaining to different tiles; RYMV
  is an example.} bond structure between those proteins.

\begin{figure}[ht]
\begin{center}
\raisebox{-1.3cm}{\includegraphics[width=10.1cm,keepaspectratio]{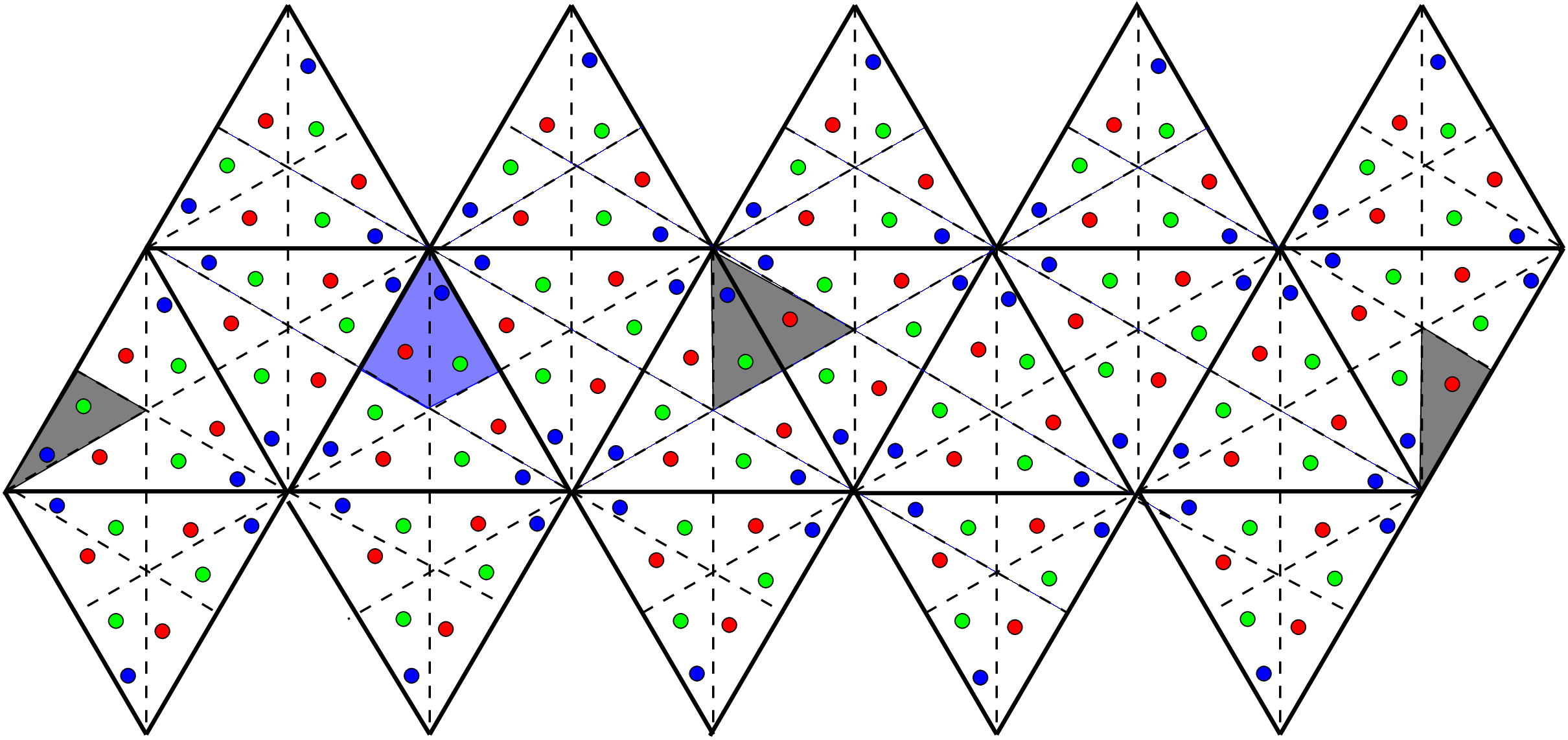}}
\end{center}
\caption{\em In the above picture, the location of the $T=3$ RYMV capsid proteins coincide with the location of the centre of mass of each of them, calculated from the experimental data collected in the file 1f2n.vdb. A triangular tiling (dashed lines) is superimposed on the icosahedral structure, and the colour coding is faithful to that of the VIPER website: the A chain is blue, the B chain, red and the C chain, green. The grey shaded triangular prototiles highlight trimer interactions between capsid proteins, while the blue shaded region corresponds to the fundamental domain of the  proper rotation subgroup ${\cal I}$ of the full icosahedral group $H_3$ . }
\label{fig:icosirymvcktiling}
\end{figure}

\begin{figure}[ht]
\begin{center}
\raisebox{-1.3cm}{\includegraphics[width=12.1cm,keepaspectratio]{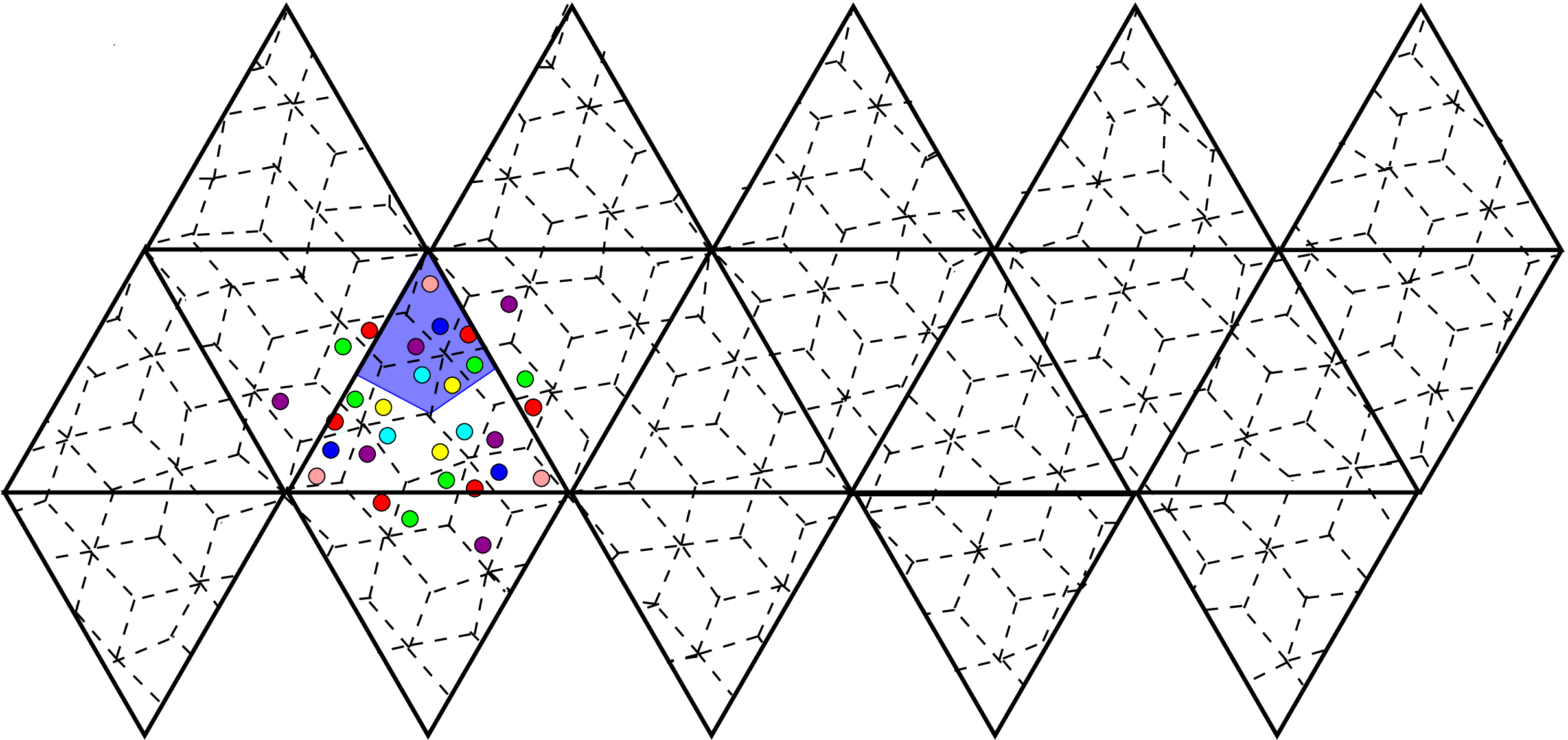}}
\end{center}
\caption{\em The location of the $T=7\ell$ HK97 capsid proteins coincide with the location of the centre of mass of each of them, calculated from the experimental data collected in the file 2fte.vdb. A rhomb tiling (dashed lines) is superimposed on the icosahedral structure, and the colour coding is faithful to that of the VIPER website: A chain (blue), B chain (red),  C chain (green), D chain (yellow), E chain (cyan), F chain (purple) and G chain (pink). The rhomb prototiles highlight four types of dimer interactions between capsid proteins, while the blue shaded region corresponds to the fundamental domain of the  proper rotation subgroup ${\cal I}$ of the full icosahedral group $H_3$ . }
\label{fig:icosihk97rhomb}
\end{figure}

The Rice Yellow Mottle Virus (RYMV) belongs to the Sobemovirus
genus. It is classified as a $T=3$ virus in the Caspar-Klug labelling
system \cite{Caspar:1962a}, and its icosahedral capsid accommodates
180 coat proteins or subunits which are clustered in 12 pentamers
around the 5-fold axes and 20 hexamers about the 3-fold global
symmetry axes of the icosahedron. The location of the proteins are
consistent with a triangular tiling \`a la Caspar-Klug, and each
triangular tile encodes trimer interactions between coat proteins, as
represented in Fig.~\ref{fig:icosirymvcktiling}.
The HK97 bacteriophage on the other hand has a $T=7\ell$ capsid made
of 420 proteins arranged in 12 pentamers and 60 hexamers, with four
types of dimer interactions modelled by rhomb prototiles; see
Fig.~\ref{fig:icosihk97rhomb}. The SV40 virus is a member of the
Polyomaviridae family and has a pseudo $T=7d$ capsid which
accommodates 360 coat proteins organised in pentamers through two
types of spherical prototiles, namely rhombs, encoding two types of
dimer interactions, and kites encoding trimer interactions, as
represented in Figure 2 of reference \cite{Keef:2005a}. SV40 is an
example of an all-pentamer capsid, for which the Caspar-Klug
classification is not applicable, and whose symmetries are 
captured by Viral Tiling Theory.

In order to extract qualitative features of vibrational patterns from
viral capsids, we restrict ourselves to a coarse-grained approximation
where each capsid protein is replaced by a point mass whose location
coincides with the centre of mass of the protein considered. This
centre of mass is calculated by taking into consideration all
crystallographically identified atoms of the protein, according to
data stored in the Protein Data Bank or equivalently the VIPER
website. We then assess how much deviation there is between the above
distribution of point masses and a theoretical distribution exhibiting
a centre of inversion. On the basis of the experimental data available
to us, we argued in \cite{ElSawy:2007a} that the SV40 capsid has an
approximate centre of inversion, while RYMV \footnote{The argument for
  RYMV is similar to the argument given for TBSV in \cite{ElSawy:2007a}.}
and HK97 do not. This has subtle consequences for the
group-theoretical properties of normal modes of vibrations: when the
capsid exhibits an effective centre of inversion, the group involved
is the full icosahedral group $H_3$ with 120 elements (usually called
${\cal I}_h$ in the science literature), while it is reduced to its
subgroup ${\cal I}$ of 60 proper rotations in the absence of a centre
of inversion.

A viral capsid with $N$ `point mass' coat proteins has $3N$ degrees of
freedom, and hence $3N$ modes of vibrations, of which 6 are associated
with 3 rotations and 3 translations of the capsid as a whole. These
are therefore not genuine normal modes of vibration. Group theory
accounts for the degeneracies of these vibrational modes, and provides
a mean to organize the normal mode spectrum of a given
capsid~\cite{Cornwell1}. A key ingredient in this exercise is the
construction of the displacement representation of the given capsid,
which is a reducible representation of $H_3$ or ${\cal I}$ according
to whether the distribution of capsid proteins exhibits a centre of
inversion or not. Such representation consists of 120 (resp. 60)
matrices $\Gamma^{\text{displ}}_{3N}(g),\,g \in H_3 \,(\rm resp.\,{\cal I}$)
of size $3N \times 3N$, which encode how proteins are interchanged
under the action of each element~$g$, as well as how the displacements
of each protein from the equilibrium position are rotated under the
action of $g$.  The latter information is gathered in $3 \times 3$
rotation matrices $R(g)$ which form an irreducible representation of
$H_3$ (resp. ${\cal I})$, while the former is encoded in permutation
matrices $P(g)$ of size $N \times N$, so that we have 
\begin{equation}
\Gamma^{\text{displ}}_{3N}(g)=P(g) \otimes R(g), \qquad \qquad \forall g \in
H_3 \,({\rm resp.}\, {\cal I}).  
\end{equation} 
The permutation matrices $P(g)$ act on vectors whose components are
the vector positions $\vec{r}^{\,\,0}_i, i=1,..,N$ of the N proteins
at equilibrium. The entry $P_{ij}(g)$ of the permutation matrix is~1
if $\vec{r}^{\,\,0}_j$ is mapped on $\vec{r}^{\,\,0}_i$ by $g$, and is
zero otherwise.

Once the displacement representation is constructed, it remains to invoke the well-known property that it can be written in block diagonal form with the help of a ($3N \times 3N$) matrix $U$~\footnote{The explicit form of the matrix $U$ is not needed at this stage, but rather when the force matrix is partially diagonalised to obtain the frequencies of vibration.} with
 \be \label{block}
U\Gamma^{\text{displ}}_{3N}(g)U^{-1}=\Gamma^{\text{displ}\,\,'}_{3N}(g)=\oplus_p n_p\Gamma^p(g),
\ee
where the multiplicities $n_p$ are obtained via the following character formula
\be \label{multiplicities}
n_p=\frac{1}{{\rm dim}\, H_3}\sum_{g \in H_3} \chi^{\text{displ}}(g)^*\,\chi^p(g)\qquad {\rm or}\qquad n_p=\frac{1}{{\rm dim\, {\cal I}}}\sum_{g \in  {\cal I} } \chi^{\text{displ}}(g)^*\,\chi^p(g).
\ee
The characters $\chi^p(g)$ of irreducible representations of the icosahedral group can be found in \cite{ElSawy:2007a}, while the characters of the displacement representations $\chi^{\text{displ}}(g)$ are obtained by inspection of the displacement representation considered. Note that, in view of the very definition of the permutation matrices $P(g)$ given in the previous subsection, and the fact that the characters of a representation are the traces of its constituent matrices, one has 
\be \label{traces}
\chi^{\text{displ}}(g)= {\rm Tr}\,(P(g))\,{\rm Tr}\,(R(g))=\pm ({\rm number\, of\, proteins\, unmoved\, by\,}g)\cdot (1+2\cos\theta),
\ee
where $\theta$ is the angle of the proper rotation associated with $g$, and the minus sign is taken when $g \in H_3 \setminus {\cal I}$. So $\chi^{\text{displ}}(g)$ is zero when $\theta=\frac{2\pi}{3}$ or whenever $g$ is such that no protein of a given capsid  is kept fixed under its action. 

The decomposition of the displacement representation of  a given capsid boils down to the knowledge of the coefficients $n_p$ in \eqref{multiplicities} which, in view of the expression \eqref{traces}, are non zero whenever at least one capsid protein is unmoved under the action of an element $g$ (and $\theta \neq \frac{2\pi}{3}$). 

It can be shown that distributions of capsid proteins with no centre of inversion are such that the only group element which keeps any `point mass' protein unmoved is the identity element $g=e$ (and under its action, all $N$ proteins are obviously fixed). The second expression in \eqref{multiplicities} thus yields
\be
n_p=\frac{1}{{\rm dim}\,{\cal I}}\chi^{\text{displ}}(e)^*\,\chi^p(e) = \frac{3N}{60}\, \chi^p(e),
\ee
where we used $\rm{dim}\, {\cal I} =60$ and $\chi^{\text{displ}}(e)= 3N$ (taking the plus sign and $\theta=0$ in \eqref{traces}). Recalling that $\chi^p(e)=p$, we arrive at the following decomposition formula,
\be \label {decomp}
 \Gamma^{\text{displ}\,'}_{3N}=\frac{3N}{60}\left\{  \Gamma^1_+ + 3\Gamma^3_+ + 3\Gamma^{3'}_+ +4\Gamma^4_+ +
5\Gamma^5_+ 
\right\}.
\ee
The number $N$ of  capsid proteins is always a multiple of sixty,  $N=60k$. In the many cases where the proteins are organised in 12 pentamers and a number of hexamers, $k$ is the $T$-number of the Caspar-Klug nomenclature. Then, the number of non-degenerate normal modes in the singlet (symmetric) representation $\Gamma^1_+$ is $3T$, while the number of $p$-fold degenerate normal modes (corresponding to the $p$-dimensional representation $\Gamma^p_+$)  is $3p^2T$, for $p=3,4$ and $5$. In particular, $N=180$ for RYMV, and the displacement representation decomposes into
 \be
 \Gamma^{\text{displ}\,'}_{540,\text{RYMV}}= 9\Gamma^1_+ + 27\Gamma^3_+ + 27\Gamma^{3'}_+ +36\Gamma^4_+ +
45\Gamma^5_+ .
\ee
The 6 non-genuine modes belong to two copies of the $\Gamma^3_+$ irreducible representation. There are nine non-degenerate and forty-five 5-fold degenerate Raman active modes, as well as  twenty-five  3-fold degenerate infrared active modes.

Since $N=420$ for HK97, the displacement representation decomposes into
 \be
 \Gamma^{\text{displ}\,'}_{1260,\text{HK97}}= 21\Gamma^1_+ + 63\Gamma^3_+ + 63\Gamma^{3'}_+ +84\Gamma^4_+ +
105\Gamma^5_+ ,
\ee
and by the same argument as above, one arrives at  twenty-one non-degenerate and one hundred and five 5-fold degenerate Raman active modes, as well as sixty-one 3-fold degenerate infrared active modes.

The normal modes of the SV40 capsid would be organised according to the decomposition \eqref{decomp} with $N=360$ if we were not taking into account that the protein distribution on the capsid exhibits an approximate centre of inversion. We would have
 \be
 \Gamma^{\text{displ}\,'}_{1080,\text{SV40}}= 18\Gamma^1_+ + 54\Gamma^3_+ + 54\Gamma^{3'}_+ +72\Gamma^4_+ +90\Gamma^5_+ .
\ee

 Instead, 
we use the first expression in \eqref{multiplicities} and note that the distribution of `point-mass' proteins on the capsid is such that, besides the identity element $g=e$ in $H_3$
which leaves all $N$ proteins unmoved, the fifteen rotations $g^{(i)}_2, i=1,..,15$ about the 2-fold axes of the icosahedron, when combined with the inversion $g_0$, produce 15
further elements $g_0g^{(i)}_2$  which altogether leave 24 capsid proteins unmoved. Those fifteen group elements are in the same conjugacy class and therefore have the same character $\chi^p(g_0g^{(i)}_2)=1$ for $p=1, 5$, $\chi^p(g_0g^{(i)}_2)=-1$ for $p=3, 3'$ and $\chi^4(g_0g^{(i)}_2)=0$. Taking into account that for these group elements, $Tr ( R (g_0g^{(i)}_2) )=-(1+2\cos \pi )=1$, we arrive at the following decomposition of the displacement representation,
\be \label{decompsv40}
\Gamma^{\text{displ}\,'}_{1080,\text{SV40}}= 12\Gamma^1_+ + 24\Gamma^3_+ + 24\Gamma^{3'}_+ +36\Gamma^4_+ +
48\Gamma^5_+ 
+ 6\Gamma^1_- + 30\Gamma^3_- + 30\Gamma^{3'}_- + 36\Gamma^4_- +
42\Gamma^5_-.
\ee
The six non-genuine modes of vibrations are confined to one copy of the 3-dimensional irreducible representation $\Gamma^3_+$, and one copy of the  3-dimensional irreducible representation $\Gamma^3_-$. There are twelve non-degenerate and  forty-eight 5-fold degenerate Raman active modes, and fifty-two 3-fold degenerate infrared active modes . 

\section{Low frequency modes}

\begin{figure}[ht]
\begin{center}
(a)\includegraphics[width=4.2cm,keepaspectratio]{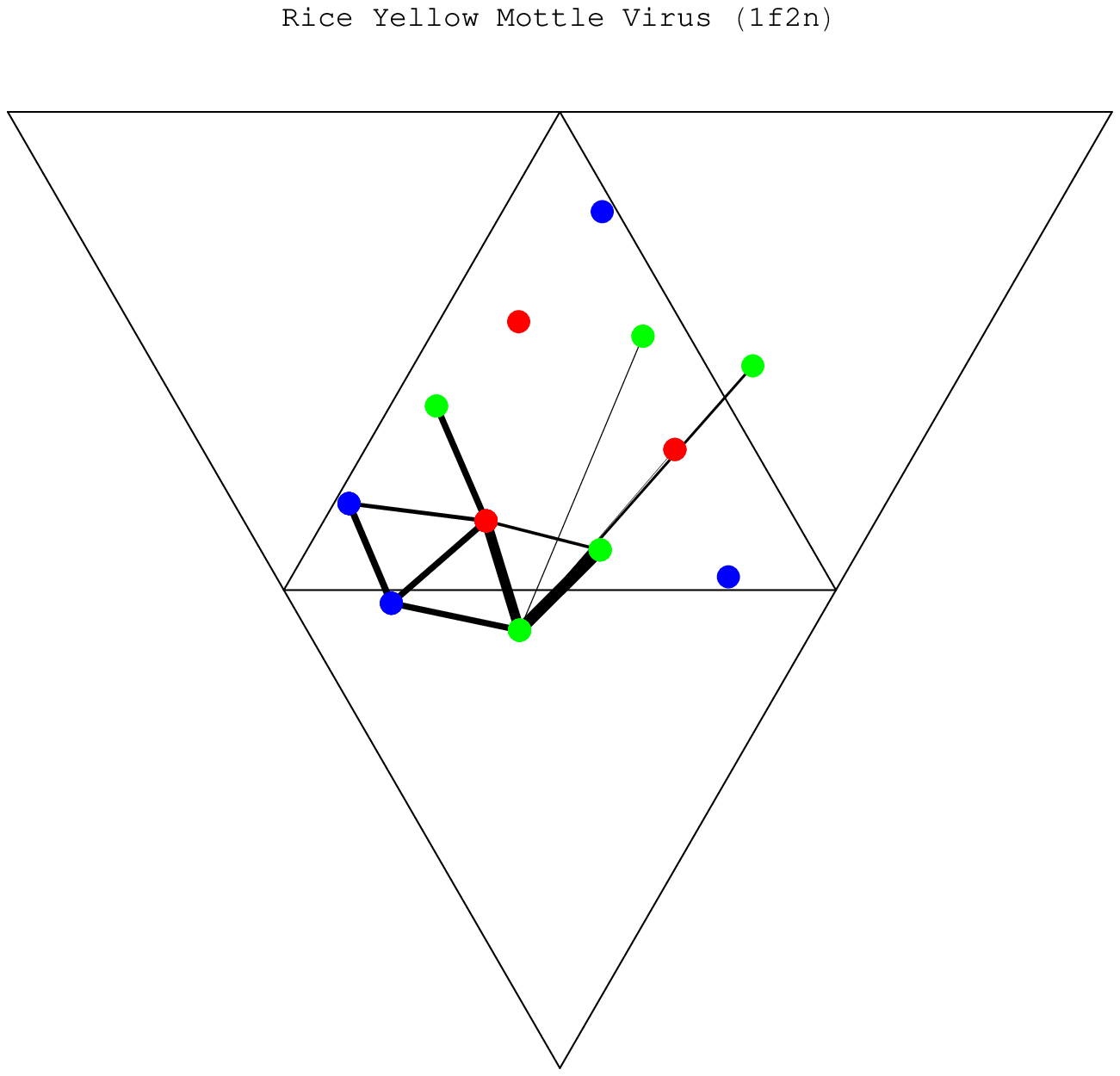}
(b)\includegraphics[width=5.2cm,keepaspectratio]{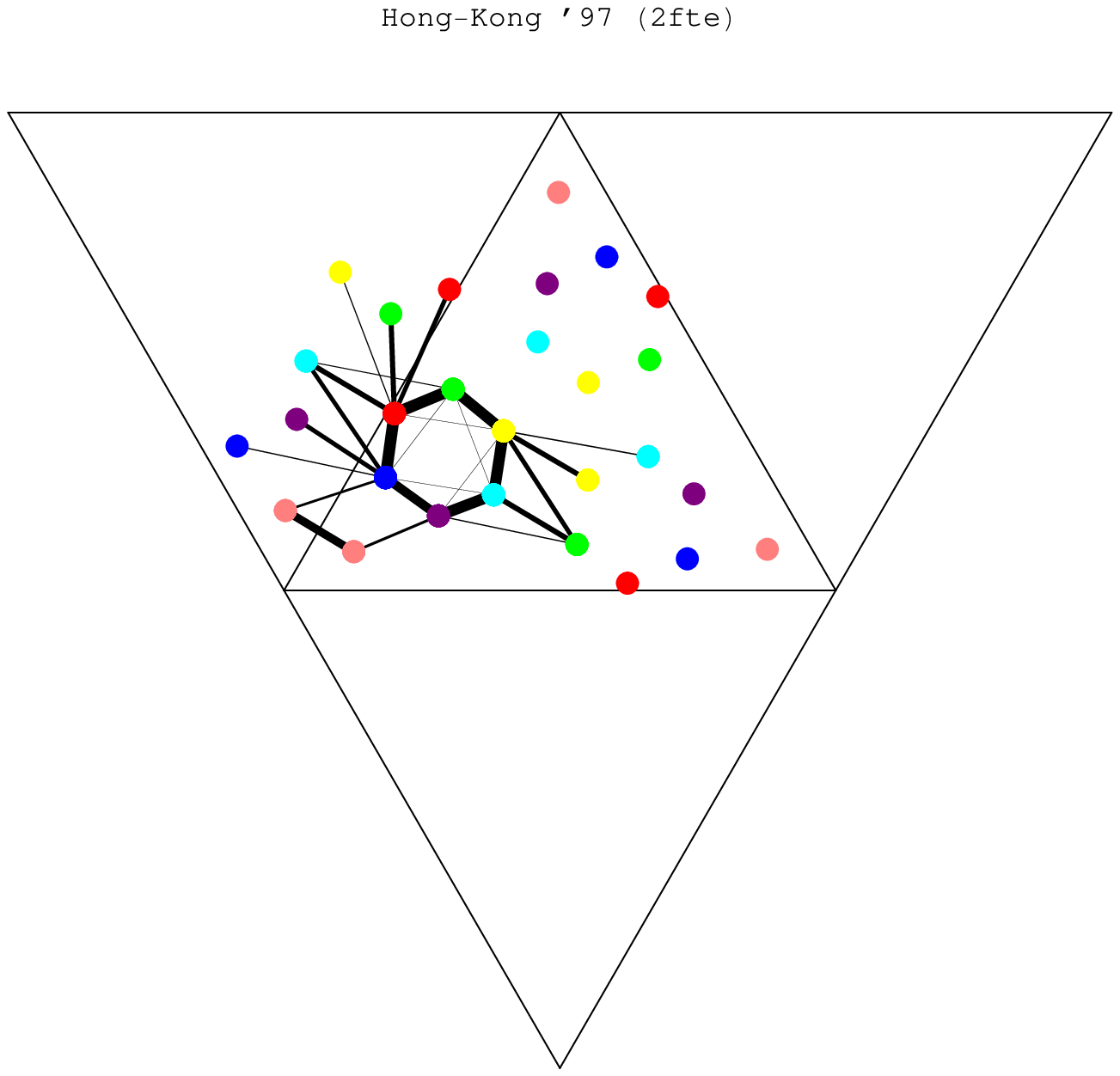}
(c)\includegraphics[width=4.2cm,keepaspectratio]{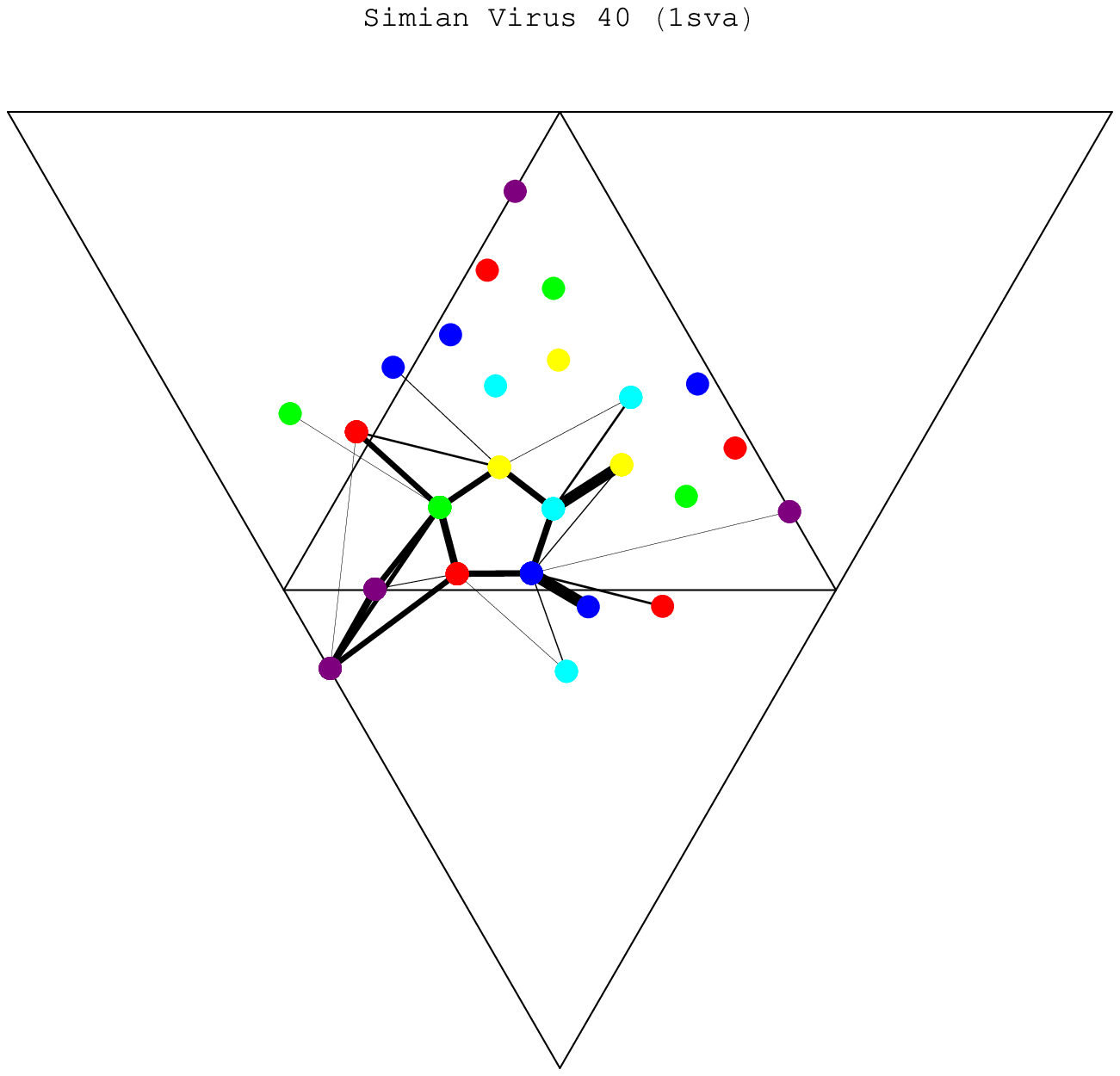}
\end{center}
\caption{\em{Inter-protein bonds are given for (a) RYMV, (b) HK97 and (c) SV40. The relative strengths are represented by line segments of varying thickness, from the strongest bonds (thick lines) to the weakest (thin lines). The above diagrammes should be read in conjunction with the figures of Section 2.}}
\label{bonds}
\end{figure}

Our calculation of the low frequency normal modes is based on a spring-mass model, where the $N$ `point-mass' proteins are connected by  a network of elastic forces described by a harmonic potential which is manifestly rotation and translation invariant,
\be 
V= \displaystyle{\sum_{\substack{m <n\\m,n =1}}^N\,\frac{1}{2} \kappa_{mn}\,(|\vec{r}_m-\vec{r}_n|-|\vec{r}^{\,0}_m-\vec{r}^{\,0}_n|)^2}.
\ee
The associated force matrix or `Hessian' is given by
\be \label{force}
F^{ij}_{mn}=\frac{\partial^2 V}{\partial r_m^i\partial r_n^j}\Bigg |_{x=x_0}=\left \{
\begin{array}{ll}
\displaystyle{ \sum_{p \neq m} \kappa_{mp}\frac{(r_m-r_p)^i\,(r_m-r_p)^j}{(r_m-r_p)^2}\Bigg  |_{r=r_0}}& {\rm if}\, m=n,\\
\displaystyle{- \kappa_{mn}\frac{(r_m-r_n)^i\,(r_m-r_n)^j}{(r_m-r_n)^2}\Bigg  |_{r=r_0}} &{\rm otherwise.}
 \end{array}\right .
\ee In the above formulae, the vector $\vec{r}^{\,0}_m$ refers to the
equilibrium position of protein $m$ and the vector $\vec{r}_m$ of
components $r_m^i$, to its position after elastic displacement, all
vectors originating at the centre of the capsid. The masses of the
proteins are all set to unity (reflecting the fact that the various
protein chains in a capsid have masses which are too good approximation
identical), and $\kappa_{mn}$ is the spring constant of the spring
connecting protein $m$ to protein $n$. The set of non-zero spring
constants we choose, i.e. the topology of the elastic network we
adopt, is dictated by the information derived from the association
energies listed in VIPER for RYMV (1f2n.vdb), HK97 (2fte.vdb) and SV40
(1sva.vdb). Fig.~\ref{bonds} encodes the bonds provided by VIPER {\em
  before} acting on them with the icosahedral group in order to
generate the complete spring network.

We have used the relative values of these energies, and therefore we
are left with one parameter $\kappa$ in the force matrix, which sets
the overall scale of the vibration frequencies. We are not aware of
any experimental measurements of association energies between capsid
proteins for the viruses and phages we are considering, and the {\em
  absolute} theoretical values calculated in \cite{Reddy:1998a} must
be taken with extreme caution.

The force matrices $F^{ij}_{mn}$ we consider here have size $3N \times
3N$ with $N=180$ for RYMV, $N=420$ for HK97 and $N=360$ for
SV40. Although computers can handle a brute force diagonalisation of
such matrices, and provide eigenvalues which are the square of the
sought frequencies of vibration of normal modes, a group theoretical
approach reduces considerably the size of the matrices to be
diagonalized and above all, yields information on the distribution of
normal modes within irreducible representations of the icosahedral
group. This proves to be useful in an analysis of universal features
of such vibrations.

\begin{figure}[ht]
\begin{center}
(a)\includegraphics[width=6.2cm,keepaspectratio]{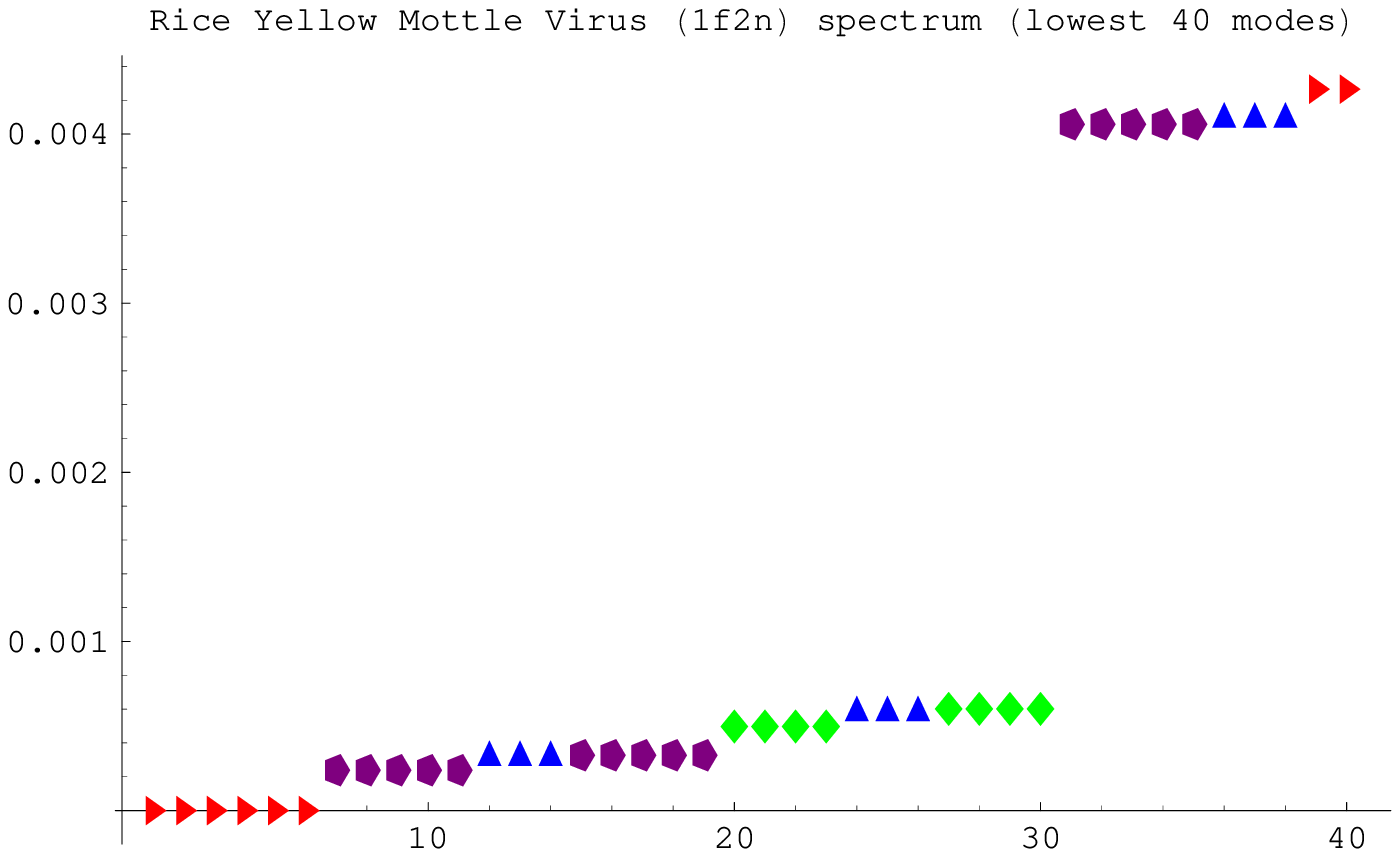}\qquad\qquad
(b)\includegraphics[width=6.4cm,keepaspectratio]{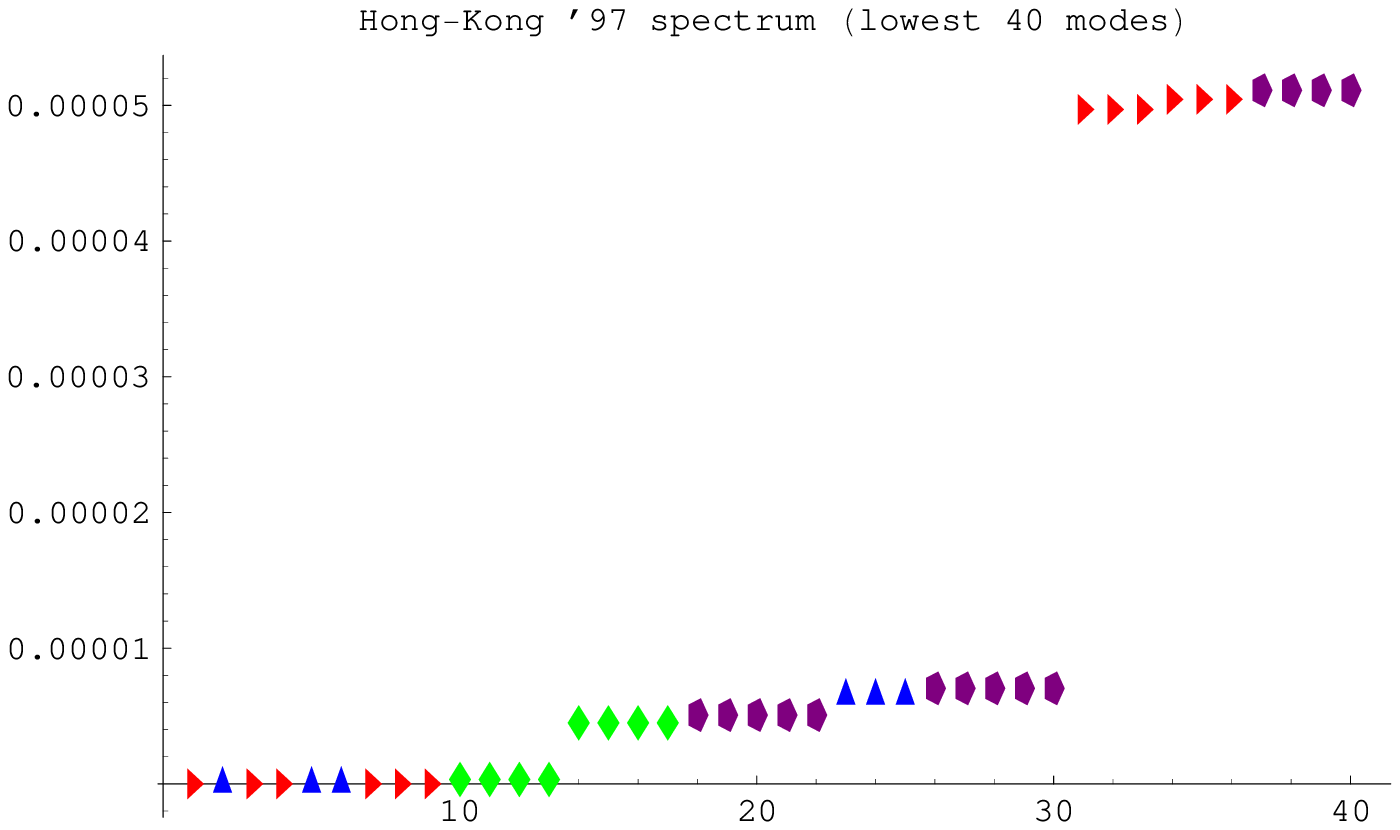}
\end{center}
\caption{\em{Spectrum of low frequency normal modes for the RYMV
    capsid (a) and the HK97 capsid (b). The triangular-shape modes
    $\rhd$ belong to  3-dimensional irreducible representations
    $\Gamma_+^3$ of the icosahedral group, while the triangular-shape
    modes $\triangle$ belong to 3-dimensional irreducible
    representations $\Gamma_+^{3'}$. Accordingly, the diamond-shape
    modes belong to 4-dimensional irreducible representations and the
    pentagon-shape modes to 5-dimensional irreducible
    representations. The $x$-axis labels the normal modes while the
    $y$-axis gives the wave numbers in cm$^{-1}$ (up to an overall
    normalisation which cannot be fixed from Viper data).}}
\label{spectra}
\end{figure}

We have calculated the lowest frequency modes of vibration for the
RYMV, HK97 and SV40 capsids using well-known group theoretical
methods. The association energies listed in Viper for RYMV allow for a
stable capsid. Crucial to the stability are the C-arms linking
together distant proteins of the C chain in Fig.~\ref{bonds}a. The
spectrum of the first 40 low frequency modes is presented in
Fig.~\ref{spectra}a. Apart from the six zero modes associated with the
rotations and translations of the capsid as a whole, and which belong
to two copies of the irreducible representation $\Gamma_+^3$ of ${\cal
  I}$, one notices a cluster of 24 normal modes of very low and
similar frequencies organized in a sum of irreducible representations
according to
$\Gamma_+^5+\Gamma_+^{3'}+\Gamma_+^5+\Gamma_+^4+\Gamma_+^{3'}+\Gamma_+^4$. This
low plateau is disrupted by a significant jump in wave number.

A similar analysis was performed for HK97. The association energies
listed in Viper for HK97 allow for a nearly-stable capsid, with nine
strictly zero modes instead of the six expected. The 21 subsequent
modes have similar frequencies, as can be seen from
Fig.~\ref{spectra}b. They are organized in the following sum of
irreducible representations of ${\cal I}$:
$\Gamma_+^4+\Gamma_+^{4}+\Gamma_+^5+\Gamma_+^{3'}+\Gamma_+^{5}$. If
the spurious triplet of zero modes were lifted by the addition of
extra bonds in the spring-mass model of HK97, one would again observe
a cluster of 24 low frequency normal modes forming a plateau disrupted
by a jump of the same scale as that appearing in RYMV. We have
observed this phenomenon in a large number of viral capsids, and we
will detail our findings in \cite{kas_virdyn}.
  
\begin{figure}[ht]
\begin{center}
\raisebox{-1.3cm}{\includegraphics[width=7.5cm,keepaspectratio]{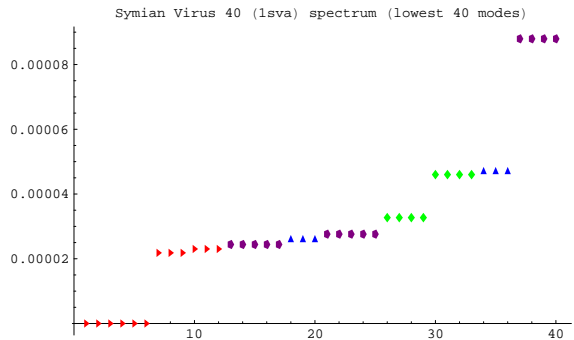}}
\end{center}
\vspace{-1ex}
\caption{\em Low frequency normal modes for the SV40 capsid. Symbol conventions as in Fig.~\ref{spectra}.}
\label{SV40spectrum}
\end{figure}

The SV40 case is particularly interesting because it does not quite
fit with the above observations. As mentioned in Section 2, the viral
capsid has a near centre of inversion, and one might want to explore
the implications of treating the normal mode analysis with a
symmetry-corrected `point-mass' protein distribution. This, however,
destabilizes the capsid, as the vertices of some triangular cells in
the network become collinear. We will therefore refrain from
considering a capsid with a centre of inversion, and perform the
normal mode analysis as in the two previous cases (RYMV and
HK97). Once more, we have plotted the low frequency spectrum in
Fig.~\ref{SV40spectrum}.

Apart from the six zero modes associated with the rotations and
translations of the capsid as a whole, one could argue that the next
23 modes should be considered as a cluster since their frequencies are
very similar. However the plateau in this case is not disrupted by a
spectacular jump in frequency, as the seven subsequent frequencies are
roughly 1.6 larger than the first 23 non-zero modes.  Early comparison
with Murine Polyomavirus vibrational patterns does not shed light on the
significance of these all-pentamer viral capsids spectra, and more
investigations are needed.

\section{Conclusion}

We have discussed the vibration spectrum of icosahedral virus capsids,
obtained from a coarse-grained model in which protein chains and their
interactions are replaced by a spring-mass model. The goal of this
programme is to understand, in a top-down approach, how properties of
the capsid structure, such as an approximate inversion symmetry or a
particular tiling type, reflect themselves in the vibrational
spectrum.  We believe this a useful complement to existing bottom-up
approaches, which are rooted in all-atom computations.

A comparison of our results with the spectra obtained in earlier
all-atom computations reveals some interesting similarities. The most
striking one is the existence of a low-frequency plateau of 24 modes,
separated by a rather large gap from the remainder of the
spectrum. This plateau is present for RYM as well as HK'97 and a large
number of other virus capsids. It has been seen before in isolated
examples~\cite{Tama:2002a,Tama:2005a,Rader:2005a}, but the simplicity of our
model offers a better chance to understand the general reason behind
its existence (more details will be provided
in~\cite{kas_virdyn}). 

While Viral Tiling Theory provides a beautiful classification of the
structure of virus capsids, its role in understanding the vibration
spectra is at present less clear. Besides the bonds which bind
together proteins on the same tile, many other bonds are required in
order to obtain a stable capsid. These other, inter-tile bonds are
often of a similar strength as the bonds on a single tile. In fact, it
is an interesting mathematical problem to understand the best network
topology (in terms of the optimal number of bonds) required for
stability of a capsid. 

The present analysis focuses exclusively on the viral
capsid, ignoring in particular the interaction of the virion with its
environment and the presence of matter within the shell, which are
undoubtedly worth considering in more elaborated models. Large-scale
simulations have revealed that some virus capsids are unstable without
RNA content~\cite{Freddolino:2006a}. It would be interesting to understand
this instability, as well as the effect of RNA content, for larger
classes of capsids.


\begingroup\raggedright\endgroup

\end{document}